# Traveling Salesman Problem solution using Magnonic Combinatorial Device

Mykhaylo Balinskiy and Alexander Khitun

*Department of Electrical and Computer Engineering, University of California -Riverside, Riverside, California, USA 92521*

Correspondence to akhitun@engr.ucr.edu

**Abstract:** Traveling Salesman Problem (TSP) is a decision-making problem that is essential for a number of practical applications. Today, this problem is solved on digital computers exploiting Boolean-type architecture by checking one by one a number of possible routes. In this work, we describe a special type of hardware for the TSP solution. It is a magnonic combinatorial device comprising magnetic and electric parts connected in the active ring circuit. There is a number of possible propagation routes in the magnetic mesh made of phase shifters, frequency filters, and attenuators. The phase shifters mimic cities in TSP while the distance between the cities is encoded in the signal attenuation. The set of frequency filters makes the waves on different frequencies propagate through the different routes. The principle of operation is based on the classical wave superposition. There is a number of waves coming in all possible routes in parallel accumulating different phase shifts and amplitude damping. However, only the wave(s) that accumulates the certain phase shift will be amplified by the electric part. The amplification comes first to the waves that possess the minimum propagation losses. It makes this type of device suitable for TSP solution, where waves are similar to the salesmen traveling in all possible routes at a time. We present the results of numerical modeling illustrating the TSP solutions for four and six cities. Also, we present experimental data for the TSP solution with four cities. The prototype device is built of commercially available components including magnetic phase shifters/filters, coaxial cables, splitters, attenuators, and a broadband amplifier. The device literally shows the shortest route between the four selected cities. There are three examples of finding the shortest route between the cities for three different sets of city-to-city distances. The ability to exploit classical wave superposition is the most appealing property of the demonstrated device. It allows us to check a number of possible routes in parallel without any time overhead. It provides a fundamental advantage over conventional digital computers in functional throughput. The estimated functional throughput of the combinatorial device for TSP with 25 cities exceeds $10^{35}$ Ops/s·m² which is above the limits of the existing supercomputers combined. Physical limits and challenges are also discussed.

I. Introduction

TSP is one of the most well-known combinatorial optimization problems [1]. It can be stated as follows: "Given a list of cities and the distances between each pair of cities, find the shortest possible route that visits each city exactly once and returns to the origin city." It is a non-deterministic polynomial-time hardness (NP-hard) problem which means that there is no guarantee to get the optimal route and no exact algorithm to solve it in polynomial time. TSP is important in a variety of practical applications including transportation[2], scheduling[3], and genomics [4]. Mathematically, it can be defined as given a set of $n$ cities, named $\{c_1, c_2, \dots, c_n\}$, and permutations $\{\sigma_1, \sigma_2, \dots, \sigma_{n!}\}$. The objective is to choose $\sigma_i$ such that the sum of all Euclidean distances between cities in a tour is minimized. TSP can be modeled as an undirected weighted graph as shown in Fig.1, where the cities are the graph's vertices, paths are the graph's edges, and a path's distance is the edge's weight. TSP can be solved by checking one by one all possible $(n-1)!/2$ possible routes. It is relatively easy to check all possible routes for a small number of cities. For example, there are $(4-1)!/2 = 3$ routes for the TSP with four cities. The number of possible routes increases proportionally to $n$ factorial which makes it difficult to solve for a large number of cities using classical computing devices.

There were several attempts to build special hardware for TSP solution [5,6]. For instance, 6-city TSP was solved using a Kohonen-type optical network[6]. However, none of these approaches resulted in a practically viable device. Recently, a variety of optimization techniques used in artificial intelligence (AI) have been applied to TSP [7]. It may significantly speed up the TSP solution but cannot provide a fundamental advantage being implemented on conventional hardware.

Here, we consider a special type of combinatorial device in which an act of computation is associated with finding a propagation route of the wave through the selected nodes in the mesh. This approach was first described in Ref. [8]. The main idea is to exploit the advantage of classical wave superposition, where a number of waves can propagate through different routes at the same time. It may be possible to amplify only those signals that propagate through the selected sites in the mesh and accumulate a certain phase shift. Spin wave (magnonic) devices are most suitable for this purpose due to the prominent phase shifts that can be achieved in micrometer-length waveguides. The rest of the work is organized as follows. In Section II, we describe the principle of operation of the magnonic combinatorial logic device for TSP. The results of numerical modeling illustrating the TSP solution for four and six cities are presented in Section III. The experimental data for the TSP solution for four cities are presented in Section IV. The Discussion and Conclusions are given in sections V, and VI, respectively.

## II.      Material Structure and Principle of Operation

The schematics of the combinatorial device for TSP with four cities are shown in Fig.2. The core of the structure is the mesh consisting of the phase shifters, attenuators, frequency filters, and power sensors. The circles marked with letters A, B, C, and D, are phase shifters representing the four cities. Each phase shifter provides a unique phase shift which is proportional to the logarithm of a prime number. For example, city A is represented by the phase shifter providing $\pi \times Log(2)$ phase shift to the propagating signal. City B is represented by the phase shifter providing $\pi \times Log(3)$ phase shift, and so on. There are two cities A, one on the left and one on the right sides of the mesh. This is the city the salesman starts from and should return to after the trip. The phase shifters are connected to each other via waveguides. There is an attenuator inserted into each waveguide. The distance between the cities is encoded into the signal attenuation. For example, 10 miles can be taken equivalent to 1 dB attenuation. There is a set of frequency bandpass filters. These filters are aimed to ensure different frequencies for different propagation routes from city A on the left to city A on the right. For example, only a signal with frequency $f_1$ can propagate through route A-B-A; only a signal with frequency $f_2$ can propagate through route A-B-C-A, etc. There is also a set of power detectors inserted into each waveguide. These detectors are aimed to provide the output: the signal propagation route. In Fig.2, the detectors are marked as color circles. The green color is to depict the energy flow above some reference value (e.g., 1 dBm). The red color is to depict no energy flow.

There is a broadband amplifier $G$, a voltage-tunable phase shifter $\Psi$, and a voltage-tunable attenuator $R$ outside the mesh. Hereafter, we refer to these elements as the "external" part. The broadband amplifier is to provide signal amplification for all signal frequencies. The voltage-tunable phase shifter and the voltage-tunable attenuator are to control the phase shift and the attenuation of the external part. The combination of the mesh with the external part constitutes a multi-path active ring circuit [8]. There is a number of possible routes for signals to propagate in the mesh. The different propagation paths are associated with different phase shift $\Delta(f_i)$ as well as the different signal attenuation $L(f_i)$, where $f_i$ is the frequency of the signal propagating through the $i$-th route. *There is a superposition of waves propagating through the different paths in the circuit*. *However, only some of the frequencies will be amplified to enable stable auto-oscillations.* There are two conditions for auto-oscillations to occur in an active ring circuit [9]:

$G \times R \times L(f_i) \geq 1,$  (1.1)

$\Psi + \Delta(f_i) = 2\pi k,$ where $k = 1,2,3, \dots$  (1.2)

where $G$ is the gain provided by the amplifier, $R$ is the signal attenuation in the electric part, $\Psi$ is the phase shift of the external electric part. The first equation (1.1) states the amplitude condition for auto-oscillations: the gain provided by the broadband amplifier should be sufficient to compensate for losses in the mesh and losses introduced by the external attenuator. The second equation states the phase condition for auto-oscillations: the total phase shift for a signal circulating through the ring should be a multiple of $2\pi$. In this case, signals come in phase every propagation round. A more rigorous description of signal propagation in active ring circuits can be found in Ref. [9].

The computational procedure for TSP is the following. The external phase shifter $\Psi$ is set up to the value

$$\Psi = 2\pi - \sum \phi, \qquad (2)$$

where the last term on the right is the sum of the phase delays for the selected cities. In our case of four cities,

$$\sum \phi = \pi[2 \times Log(2) + Log(3) + Log(5) + Log(7)]. \qquad (3)$$

Only signals propagating through the selected cities (e.g., A,B,C,D, and A) will be amplified as the total phase shift through the ring is $2\pi$. A more detailed explanation of the selective signal amplification in a multi-path active ring circuit can be found in Ref. [8]. All other signal propagating on other paths (e.g., A-B-A, A-C-D-A, etc.) won't be amplified as the phase condition (1.2) is not satisfied.

Then, it is possible to find the shortest path (i.e., the path with minimum losses) using the external attenuator $R$. There may be a number of routes for which the total phase shift satisfies Eq.(1.1). The number of routes satisfying the amplitude condition Eq.(1.2) decreases with the increase of the external damping $R$. The shortest path will disappear the last as it may sustain the maximum external damping. There are two important points we want to emphasize regarding the principle of operation. *The system starts with a superposition of all possible frequencies propagating through all possible paths. We can amplify only the signals coming through the selected nodes/cities using the external phase shifter to satisfy the phase condition (1.2). The shortest route is found then by introducing additional damping so only the signal with minimum losses can satisfy the amplitude condition (1.1).* In the next section, we present the results of numerical modeling illustrating the TSP solution.

III. **Numerical modeling**

To illustrate the solution procedure described above, we present the results of numerical modeling for the four-cities TSP. In Fig.3(a), there is shown a map of the 20 most fun cities in U.S. (source WalletHub). The Traveling Salesman starts from Los Angeles and has to visit Miami,

Washington, Chicago, and come back to Los Angeles. The distances between the cities are taken from the Google Map application. The problem is solved using the combinatorial device shown in Fig.2. The distances between the cities are converted into signal attenuation (e.g., 1000 miles = 10 dB attenuation).

There are 27 possible routes between the four cities (e.g., Los Angles -Washington – Los Angeles, - Miami-Washington - Los Angeles). In Fig. 2(b), there are shown the travel distances for all routes. The routes are marked by #1, #2, … #27. Each of the routes is associated with a continuous sinusoidal signal of some frequency $f_1, f_2, … f_{27}$. The values of these frequencies are of no importance. We assume that the phase shifts provided by the cities do not depend on the signal frequency. The problem is solved in two steps. In Step 1, we set up the external phase shifter according to Eqs. (2) and (3). Only the routes coming through all four cities and coming back to Los Angeles will be amplified in the active ring circuit. In Fig.2(c), there are shown phase shifts for all possible routes. Only 6 routes out of 27 satisfy the phase condition (1.2) and will be amplified. In Step 2, we introduce additional damping for the external electric circuit so only the route with minimum losses can satisfy condition (1.2). In Fig.3(d), there are shown the results of numerical modeling depicting the shortest route(s) out of six possible from Step 1. Actually, there are always two routes with the shortest distance. In our case, these are Los Angeles – Miami – Washington – Chicago – Los Angeles and Los Angeles – Chicago – Washington – Miami – Los Angeles.

The above-described algorithm can be extended to a larger number of cities. For example, there are shown the schematics of a device for TSP with six cities in Fig.4. There are two more phase shifters marked as E and F added to the mesh. These shifters provide $\pi \times Log(11)$ and $\pi \times Log(13)$ phase shifts, respectively. There are more interconnects between the sites of the mesh. Each interconnect is a waveguide that includes an attenuator, a frequency filter, and a power detector. For simplicity, attenuators, filters, and detectors are not shown in Fig.4. In this example, the traveling salesperson starting from Los Angeles should visit Las Vegas, Chicago, Washington, Miami, San-Francisco, and return to Los Angeles. The map with the cities is shown in Fig.5(a). The table with city-to-city distances can be found in the Supplementary materials. There are more than 3000 possible routes as shown in Fig.5(b). Not all of these routes are coming through all the selected cities. In Step 1, the external phase shifter is tuned to

$$\Psi = 2\pi - \pi[2 \times Log(2) + Log(3) + Log(5) + Log(7) + Log(11) + Log(13)]. \qquad (5)$$

Only routes coming through all cities (i.e., the routes accumulating the required phase shift) will be amplified. The results of numerical modeling in Fig.5(c) show the routes satisfying Eq.(1.1). In Step 2, additional damping is introduced by the external attenuator. In Fig.3(d), there are shown the shortest routes (i.e., #320 and #3062) that correspond to Los Angeles – Miami – Washington – Chicago – Las Vegas – San Francisco – Los Angeles, and the route with the backward city order: Los Angeles– San Francisco – Las Vegas – Chicago – Washington – Miami– Los Angeles.

It should be noted that in the presented above examples TSP was solved using a general type computer. All possible routes were calculated one by one. Routes with a phase shift satisfying Eq.(1.1) were selected then. The shortest route out of the routes with the right phase was found by checking all the routes with the right phase. The use of conventional hardware does not provide any advantage over the existing algorithms. To speed up the TSP solution, one needs a device that utilized wave superposition and checks all possible routes in parallel.

### IV. Experimental Data

In this part, we present experimental data showing TSP solution for four cities. There is a variety of possible approaches to the practical implementation of the device shown in Fig.2. It may be a superposition of mechanical, acoustic, or electromagnetic waves propagating through all possible routes in the mesh. Our approach is aimed to explore the unique combination of properties inherent in spin waves, where magnetic waveguides can serve as phase shifters and frequency filters at the same time. Spin wave is a collective oscillation of spins in a spin lattice around the direction of magnetization. Spin waves appear in magnetically ordered structures, and a quantum of spin wave is called a "magnon"[10]. Active ring circuits with spin wave delay lines are widely used as tunable microwave sources[11]. Radiofrequency spin waves propagate in magnetic waveguides much slower compared to electromagnetic waves in coaxial cables. For instance, the group velocity of magnetostatic spin waves in a single-crystal yttrium iron garnet $Y_3Fe_2(FeO_4)_3$ (YIG) film of thickness 7 μm is about $10^4$ m/s [12]. It makes it possible to obtain large (e.g., up to 2pi) phase shifts for the propagating signal in millimeter-length waveguides. At the same time, spin wave dispersion strongly depends on waveguide magnetization. Spin waves propagating along the direction of magnetization (so-called backward volume magnetostatic spin waves (BVMSW) and spin waves propagating perpendicular to the direction of magnetization (so-called magnetostatic surface spin waves MSSW) possess different dispersion [13]. Thus, both the phase shift and the frequency window for spin wave propagation can be controlled by the locally applied magnetic field. It may be just a micro-scale magnet placed on the top of the YIG waveguide. The first proof-of-the-concept experiment on signal re-direction in the two-path combinatorial logic device was accomplished using a YIG-based active ring circuit with BVMSW and MSSW [8].

In this work, the mesh is built using commercially available YIG-based frequency filters produced by Micro Lambda Wireless, Inc, model MLFD-40540. The experimental data on the filter transmission and phase delay can be found in the supplementary materials. The schematics of the device are shown in Fig.6. In Fig.6(A), there is shown the general view of the device similar to the one in Fig.2. The main difference is in the number of routes connecting the A city on the left and the A city on the right. There are shown nine band-pass filters. The central frequencies of the filters are setup to transmit three selected frequencies: $f_1 = 2.441$ GHz, $f_2 = 2.514$ GHz, and $f_3 = 2.595$ GHz. The filters are assembled to provide three propagation routes: $f_1$ for A-B-C-D-A, $f_2$ for A-C-D-B-A, and $f_3$ for A-D-B-C-A. The device is aimed to find the shortest route depending on the set of attenuators. In Fig.6(B), there are shown the connection schematics of the prototype device. There are six double-channel bandpass frequency filters connected via coaxial cables. The two letters (e.g., AB, CD, etc.) depict the path at which the filter is introduced. For example, AB means that the filter is introduced to the path connecting cities A and B. Each of the filters introduces a phase shift to the propagating signal. This fact is exploited to minimize the number of components in the circuit The phase delay for each city is introduced by the frequency filters. The system was calibrated so the signals propagating through the three routes accumulate the same phase shifts. The amplification in the electric part is provided by three amplifiers (Mini-Circuits, model ZX60-83LN-S+) connected in series. The total amplification is about 15 dB in all experiments. There is external phase shifter (ARRA, model 9418A) which adjusted to provide the phase shift of $\pi/6$. The phase condition (1.2) is satisfied for all three routes. Experimental data on signal propagation can be found in the supplementary materials. There are 12 attenuators in the circuit, where the attenuation in dB is proportional to the distance between the cities. The power flow is detected using a unidirectional coupler (KRYTAR, model 1850) which is attached to the 12 selected places as shown in Fig.6(B). These places are marked with numbers 1,2,3 and 4. The marks have different colors (i.e., red, green, and blue) to separate signals propagating on frequencies $f_1$, $f_2$, and $f_3$, respectively. The power detected at port 1 corresponds to the input power. The power measured at ports 2,3, and 4 corresponds to the signal after traveling the first, the second, and the third cities, respectively. Also, the frequency spectra of the auto-oscillation in the active ring circuit can be measured using the Spectrum Analyzer and PNA as shown in Fig.6(B). The spectra can be found in the Supplementary materials.

In the next three examples, we present experimental data demonstrating the shortest route finding depending on the set of attenuators for different paths. It should be noted that not the bandpass frequencies nor connectivity nor the position of the external phase shifter are changed during the experiments.

Experiment 1: In Table 1.1, there are shown the values of attenuators for different paths. These values are randomly chosen based on the available components. As soon as the magnetic and electric parts are connected, the device starts to search for the resonant path. It takes less than 100 $\mu s$ for coming to the stable regime of auto-oscillations. In Table 1.2, there are shown experimental data on the power flow in the circuit. There is about the same input power for all three possible routes. Most of the power flows through the route A-C-D-B-A. There is more than a 30 dBm difference with the other two routes (i.e., A-B-C-D-A and A-D-B-C-A), which are not in resonance with the electric part. The propagation route is illustrated in Fig.7. All measurements are done at room temperature.

Experiment 2: Next, the set of attenuators was changed. In Table 2.1, there are shown the values of attenuators for different paths. In Table 2.2, there are shown experimental data on the power flow in the circuit. There is about the same input power for all three possible routes. Most of the power flows through the route A-D-B-C-A. As in the previous example, there is more than a 30 dBm difference with the other two routes (i.e., A-B-C-D-A and A-C-D-B-A). The propagation route is illustrated in Fig.8.

Experiment 3: In Table 3.1, there is shown a different set of values of attenuation for different paths. In Table 3.2, there are shown experimental data on the power flow in the circuit. Most of the power flows through the route A-B-C-D-A. As in the previous example, there is more than a 30 dBm difference with the other two routes (i.e., A-B-C-D-A and A-C-D-B-A). The propagation route is illustrated in Fig.9.

There is an infinite number of sets for attenuators mimicking the traveling distance between the four cities. In all cases, the device will find the shortest route through all the cities.

## V. Discussion

There are several observations we would like to make based on the obtained experimental data. The prototype magnonic combinatorial device does show the shortest route of propagation through the selected sites in the mesh made of frequency filters, phase shifters, and attenuators. It is critically important that the change in the signal propagation route demonstrated in the examples is only due to the change in the path attenuation. Thus, the device can be utilized for all possible combinations of path attenuation (distances) between the nodes (cities).

There is a prominent difference in the power flow through the different routes which exceeds 30 dBm at room temperature. The active ring circuit device selectively amplifies only the routes which satisfy resonant conditions (1.1) and (1.2). The phase condition (1.2) allows us to extract the routes with desired accumulated phase change. The amplitude condition (1.1) makes it possible to select only the shortest route out of many with the same accumulated phase. It takes a special step in numerical modeling for extracting the shortest route by increasing the external attenuation $R$. The extraction of the shortest route *happens naturally* in the prototype device due to the non-linear phenomena. This effect is known as mode suppression [14,15] which is observed in mechanical, acoustic, and optical resonators. It explains the energy re-distribution of power between the modes, where the mode with minimum attenuation receives the most of the amplification. It is one rare case when the experimental prototype works more efficiently than the theoretical device for numerical modeling.

It should be noted the difference between the general-type device shown in Fig.2 and the prototype shown in Fig.4. It is assumed that all routes connecting the A city on the left and the A city on the right are available in the general-type device. As mentioned before, there are "parasitic" routes like A-B-A, A-C-B-A which do not go through all the cities. The "right" routes are selected using the external phase shifter. The device shown in Fig.2 can be used for finding any route (e.g., A-B-A, A-C-D-A) by adjusting the external phase shifter. In contrast, there are only three routes (i.e., A-B-C-D-A, A-B-D-C-A, and A-C-B-D-A) in the device shown in Fig.4. It is also possible to include "parasitic" routes in the cost of additional frequency filters. For example, one would need to include two additional frequency filters tuned to the central frequency $f_4$ to have route A-B-A.

There is a variety of approaches to practical implementation of combinatorial devices. Magnonic approach has certain advantages (e.g., prominent phase shift for RF signal in micro-meter scale waveguides) and disadvantages (e.g., prominent spin wave damping). It may be possible to build optical combinatorial devices with low damping and fast signal propagation if the prominent phase delay can be achieved. *The ability to exploit classical wave superposition provides a fundamental advantage over conventional digital computers in functional throughput.* Functional throughput is a commonly accepted metric for logic devices evaluation [16], which can be defined as follows:

$$Functional\ throughput = \frac{number\ of\ operations}{area \times time}. \tag{6}$$

Conventional logic devices based on Complementary Metal-Oxide Semiconductor (CMOS) technology possess deep sub-micrometer characteristic size (e.g., 7 nm channel length) and perform one operation in a very short time (e.g., a fraction of nanosecond)[17]. However, these Boolean-type devices can accomplish only one operation at a time. In contrast, combinatorial logic devices are efficient in parallel search. This search through multiple routes is equivalent to the number of subsequent operations for the digital computer. The number of possible routes in TSP scales proportional to $(n-1)!/2$. The time of computation scales proportionally to $n \cdot l/v_g$, where $l$ is the length of the waveguide connecting the nodes in the mesh, $v_g$ is the group velocity of the signal. The area of the mesh scales proportionally to $\sqrt{n} \cdot l^2$. The functional throughput of the combinatorial devices for TSP can be estimated as follows:

$$Functional\ throughput\ (combinatorial) = \frac{n!}{(n \cdot l/v_g) \times (\sqrt{n} \cdot l^2)}. \tag{7}$$

Taking the length of the waveguide to be 10 $mm$ and the group velocity to be $10^4$ m/s, the functional throughput is skyrocketing to $10^{35}$ Ops/s·m² which is above the limits of the existing supercomputers combined.

Surprisingly, there is a lot of similarity in solving different NP-hard problems such as the Bridges of Konigsberg, TSP, and prime factorization. The use of prime numbers for marking unique mesh nodes (i.e., cities in TSP) looks very efficient and guarantees the uniqueness for the phase accumulated through the different routes. In turn, the same device can be used for prime factorization. The external phase shifter can be set to $\Psi = 2\pi - \pi Log(X)$, where $X$ is the number to be factorized. The device will find a route at which the accumulated phase matches the external phase. The examples of prime factorization with the combinatorial logic device are presented in the preceding work [8]. In the future, it may be possible to build a universal combinatorial device capable of solving a variety of NP-hard problems.

The key practical question is related to the number of parts (i.e., waveguides, frequency filters, attenuators, power detectors) for building a device for TSP with a large number of cities. In Table 4, there are shown the estimates for the number of parts for TSP with $n$ cities. The number of phase shifters for cities scales as $n + 1$. One extra phase shifter is needed for the second A city. The number of waveguides connecting the phase shifters is given by

$$\# waveguides = \sum_{i=2}^{n}(i-1) + (n-1) = \frac{n(n-1)}{2} + (n-1) = (n+2)(n-1)/2, \qquad (8)$$

where the first term on the right accounts for the number of paths in the classical TSP as shown in Fig.2. The second term on the right accounts for the additional $(n-1)$ waveguides for the second city A. There is the same number of attenuators and power detectors required for TSP with $n$ cities. A more complicated is situation with frequency filters. There are 12 single-frequency bandpass filters (4 filters for 3 routes) shown in Fig.6(A). In the prototype, we used separate cables and frequency filters to make two propagation paths between some cities (e.g., BC and DC). It would take an enormous amount $(n-1)!$ of single-frequency bandpass filters for TSP with $n$ cities following this approach. In general, an $i-th$ route in the combinatorial device should be associated with some frequency $f_i$. It leads to the grand challenge of the presented approach as the number of possible routes increases factorial with the number of cities. The number of distinct frequencies can be quite large (e.g., 1M frequencies in the frequency range from 0.5 GHz to 10 GHz with the inter-frequency separation of 0.5 MHz). However, it is sufficient for solving TSP with just 6-7 cities. In part, this problem can be resolved by utilizing the same frequency for not-crossing routes. The problem can be resolved completely by utilizing *combinatorial filters* transmitting signals comprising waves with a certain *combination of frequencies* (e.g., $\{f_1, f_2, f_5, f_{11}\}$ or $\{f_2, f_4, f_7, f_{19}\}$, etc.). In this scenario, the number of filters is the same as the number of waveguides given by Eq.(8), where each filter has a unique combination of the bandpass frequencies. In all cases, there is only one external broadband amplifier, one external tunable phase shifter, and one external tunable attenuator.

Another practical challenge is associated with the accuracy of the parts with respect to the number of cities. Less accuracy is required for the power detectors. A prominent difference in the power flow (e.g., a 20 dB difference between the active and passive routes) can be achieved regardless of the number of cities. More strict accuracy requirements are applied to the external phase shifter. Indeed, the increase of possible propagation routes inevitably decreases the difference in the accumulated phase for different routes. Also, there will be plenty of routes with the same phase shift. It may be possible to set a prominent phase difference between the routes coming through the selected nodes in the mesh by increasing the phase difference between the phase shifters. This and many other questions deserve special consideration.

## VI.    Conclusions

We described an approach to TSP solution using combinatorial devices. The principle of operation is based on the classical wave superposition, where each wave corresponds to a traveling salesman. The waves propagate through a mesh of waveguides, where phase shifters represent the cities in TSP and the distance between the cities is encoded in the signal attenuation. Only waves coming through the selected cities and accumulating the specific phase shift are amplified. The most amplified is the wave traveling through the route with minimum attenuation. TSP solution is illustrated by numerical modeling and experimentally demonstrated. The first prototype is a multi-path magnonic active ring circuit comprising magnetic phase shifters, frequency filters, and attenuators. The signal propagation route is detected via the power detector. We presented thee examples for three different sets of attenuators/city-to-city distances. The device literally finds the shortest propagating route through the cities. The operation is robust as the power difference between the active and passive routes exceeds 30 dBm. All experiments are accomplished at room temperature. This work is a step toward combinatorial logic devices for special task data processing. Potentially, combinatorial devices may provide a fundamental advantage in functional throughput compared to conventional digital devices. The ability to use classical superposition of waves is the key advantage and the most appealing property of the presented approach.

**Competing financial interests**

The authors declare no competing financial interests.

**Data availability**

All data generated or analyzed during this study are included in this published article.

**Acknowledgment**

This work of M Balinskiy and A. Khitun was supported in part by the INTEL CORPORATION, under Award #008635, Project director Dr. D. E. Nikonov, and by the National Science Foundation (NSF) under Award # 2006290, Program Officer Dr. S. Basu. The authors would like to thank Dr. D. E. Nikonov for the valuable discussions.

**Figure Captions**

**Figure 1:** Undirected weighted graph for TSP with four cities. The cities are the graph's vertices, paths are the graph's edges, and a path's distance is the edge's weight.

**Figure 2:** Schematics of the combinatorial device for TSP with four cities. The core of the structure is the mesh consisting of the phase shifters, attenuators, frequency filters, and power sensors. The circles marked with letters A, B, C, and D, are phase shifters representing the four cities. Each phase shifter provides a unique phase shift which is proportional to the logarithm of a prime number. For example, city A is represented by the phase shifter providing $\pi \times Log(2)$ phase shift to the propagating signal. City B is represented by the phase shifter providing $\pi \times Log(3)$ phase shift, and so on. There are two cities A, one of the left and one of the right sides of the mesh. This is the city the salesman starts from and should return to after the trip. The phase shifters are connected to each via waveguides. There is an attenuator inserted between each pair of cities. The distance between the cities is encoded into the signal attenuation. For example, 10 miles can be taken equivalent to 1 dB attenuation. There is a set of frequency bandpass filters. These filters are aimed to ensure different frequencies for different propagation routes from the city A on the left to city A on the right.

**Figure 3:** (A) U.S. map with the 20 most fun cities. The Traveling Salesman starts from Los Angeles and has to visit Miami, Washington, Chicago and come back to Los Angeles. (B) Results of numerical modeling showing the distances for all possible routes. The routes are marked by #1, #2, ... #27. Each of the routes is associated with a continuous sinusoidal signal of some frequency $f_1, f_2, \ldots f_{27}$. (C) Results of numerical modeling showing phase shifts for all possible routes. Only 6 routes out of 27 satisfy the phase condition (1.2) and will be amplified. (D) Results of numerical modeling depicting the shortest route(s): Los Angeles – Miami – Washington – Chicago – Los Angeles and Los Angeles – Chicago – Washington – Miami – Los Angeles.

**Figure 4:** Schematics of the combinatorial device for TSP with six cities. There are two more phase shifters compared to Fig.2 marked as E and F added to the mesh. These shifters provide $\pi \times Log(11)$ and $\pi \times Log(13)$ phase shifts, respectively.

**Figure 5:** (A) U.S. map with the 20 most fun cities. The traveling salesperson starting from Los Angeles should visit Las Vegas, Chicago, Washington, Miami, San Francisco, and return to Los Angeles. (B) Results of numerical modeling showing the distances for all possible routes. (C) Results of numerical modeling showing phase shifts for all possible routes. (D) Results of numerical modeling depicting the shortest route(s): Los Angeles – Miami – Washington – Chicago – Las Vegas – San Francisco – Los Angeles, and the backward route: Los Angeles– San Francisco – Las Vegas – Chicago – Washington – Miami– Los Angeles.

**Figure 6:** (A) General view of the prototype device for TSP with four cities. There are 12 single-frequency bandpass filters included in the scheme. The central frequencies of the filters are setup to transmit three selected frequencies: $f_1 = 2.441$ GHz, $f_2 = 2.514$ GHz, and $f_3 = 2.595$ GHz. The filters are assembled to provide three propagation routes: $f_1$ for A-B-C-D-A, $f_2$ for A-C-D-B-A, and $f_3$ for A-D-B-C-A. (B) Connection schematics of the prototype device. The phase delay for each city is introduced by the frequency filters. The system was calibrated so the signals

propagating through the three routes accumulate the same phase shifts. There are 12 attenuators in the circuit, where the attenuation in dB is proportional to the distance between the cities. The power flow is detected using a unidirectional coupler (model) which is attached to the 12 selected places marked with numbers 1,2,3, and 4. The marks have different colors (i.e., red, green, and blue) to separate signals propagating on frequencies $f_1$, $f_2$, and $f_3$, respectively.

**Figure 7: (A)** Table 1.1: Node-to-node attenuation. (B) Table 1.2: Experimental data on the power flow in the mesh. (C) The shortest route ACDBA is depicted by the power sensors.

**Figure 8:** (A) Table 2.1: Node-to-node attenuation for Example 2. (B) Table 2.2: Experimental data on the power flow in the mesh. (C) The shortest route ADBCA is depicted by the power sensors.

**Figure 9:** (A) Table 3.1: Node-to-node attenuation for Example 2. (B) Table 3.2: Experimental data on the power flow in the mesh. (C) The shortest route ABCDA is depicted by the power sensors.

**Figure 10:** Table 4.: Number of components for the combinatorial device for TSP with $n$ cities.

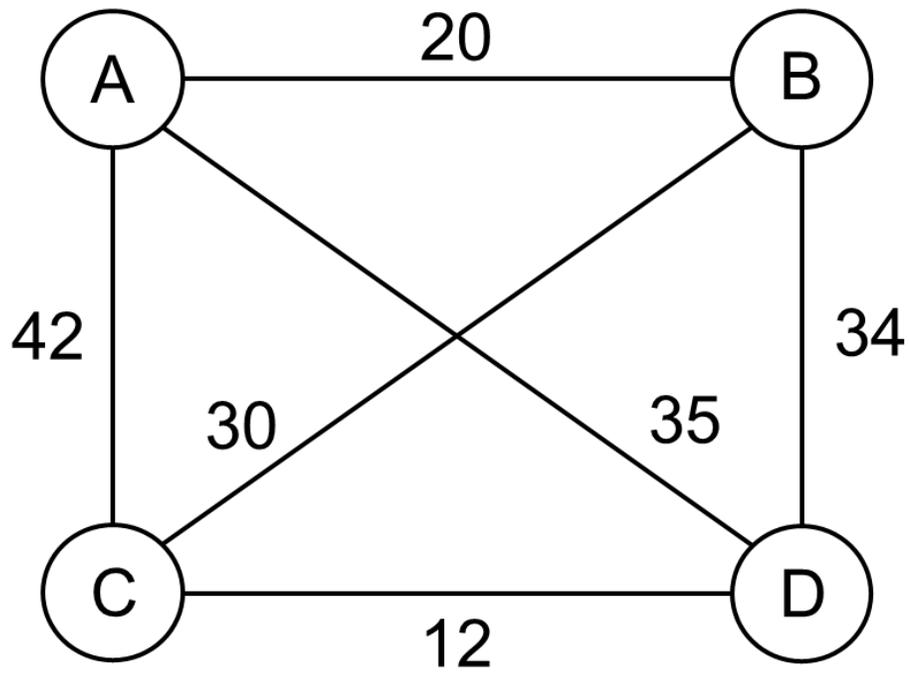

**Figure 1**

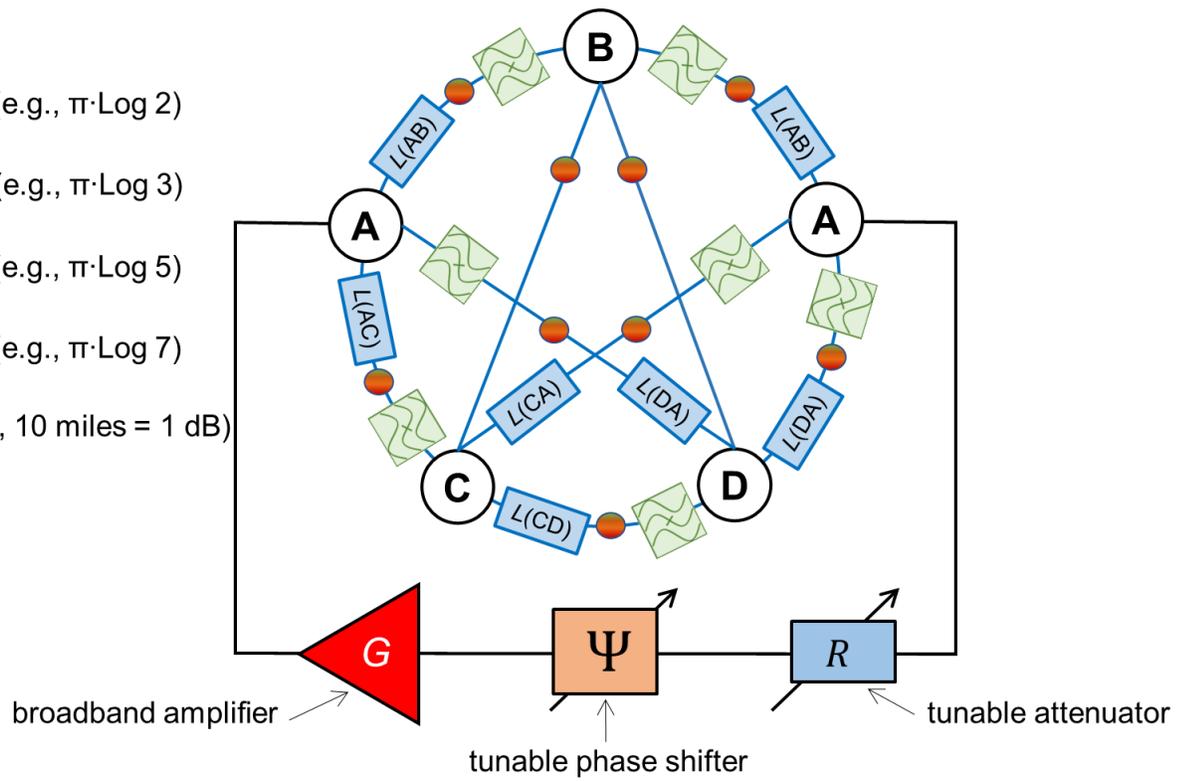

**Figure 2**

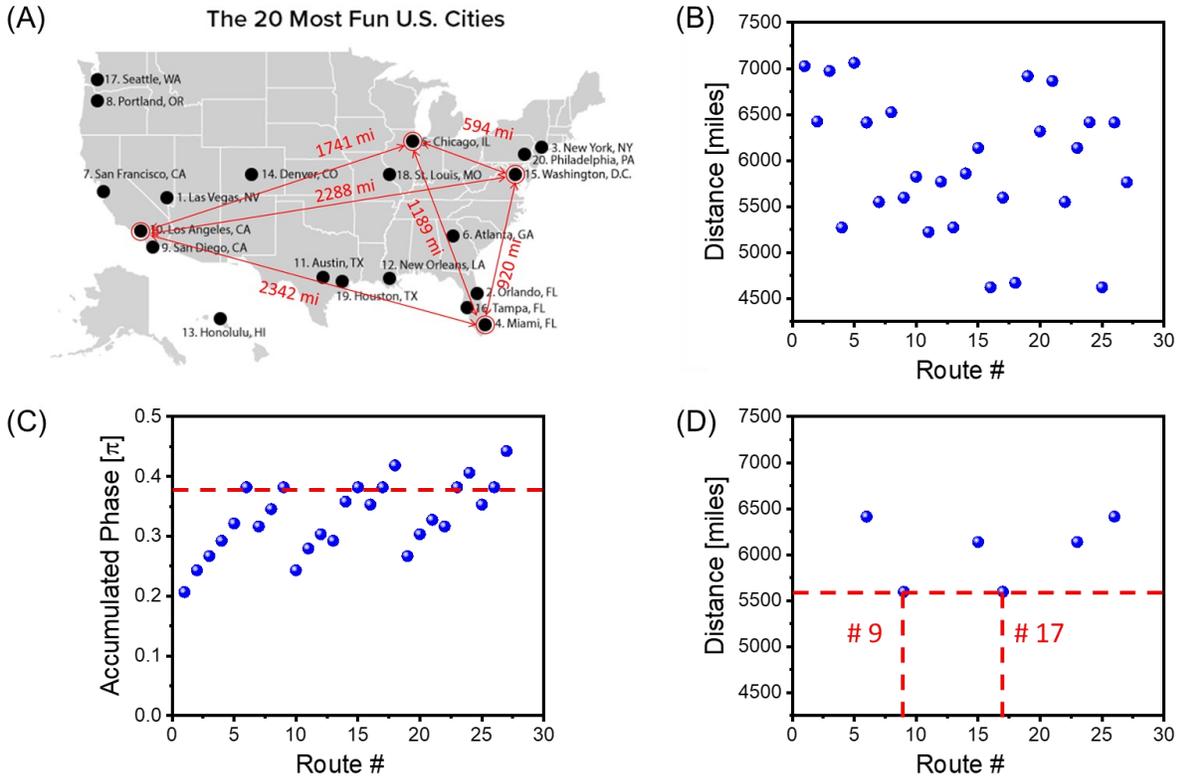

**Figure 3**

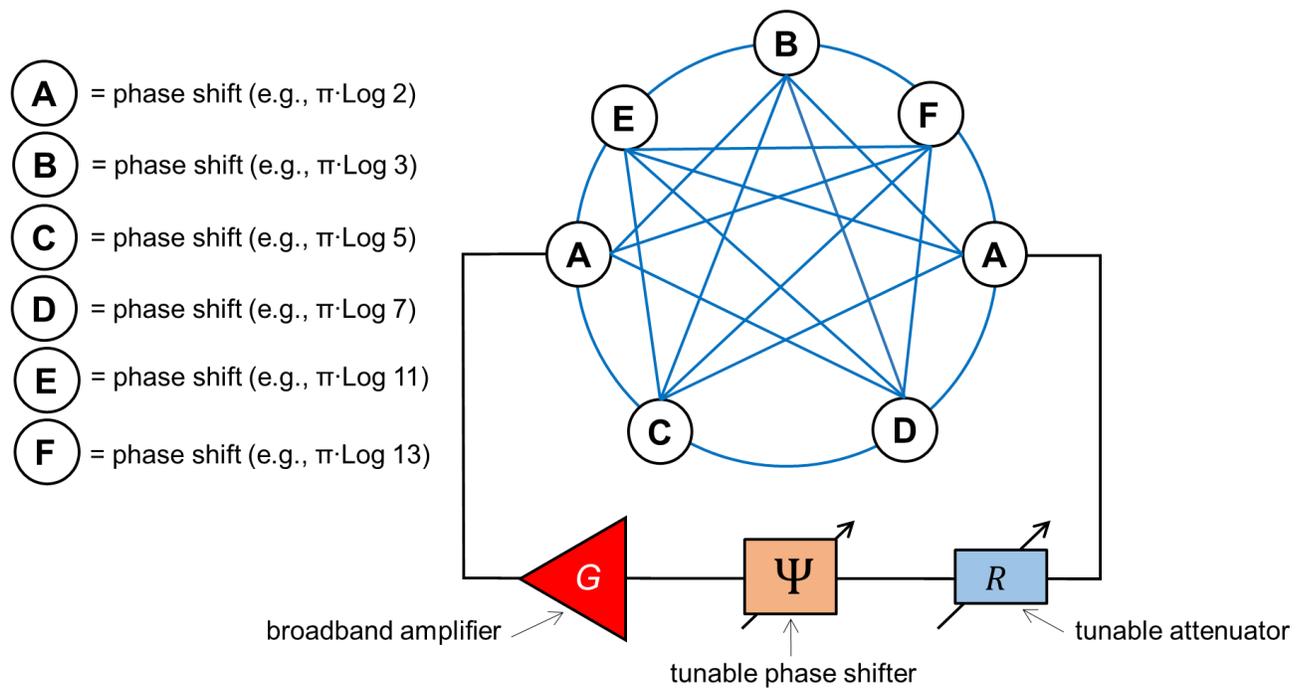

**Figure 4**

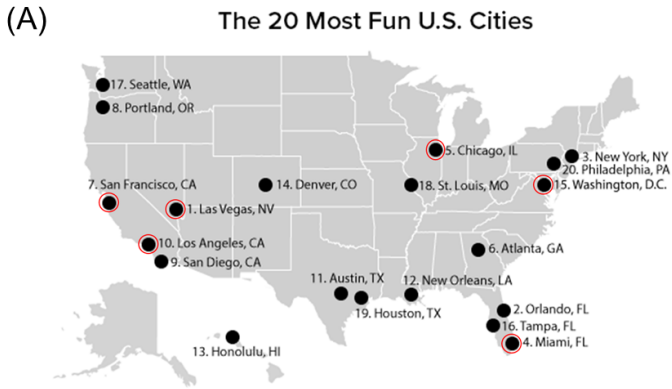
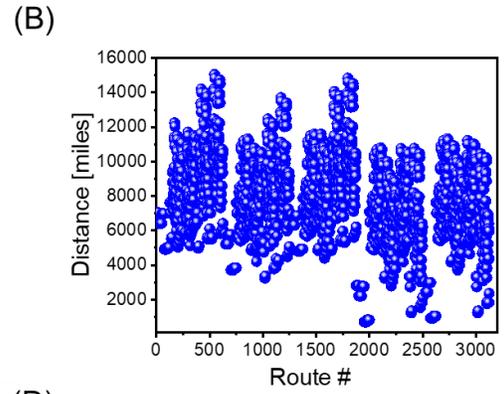
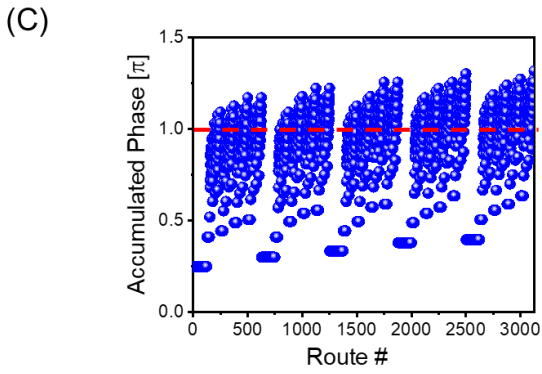
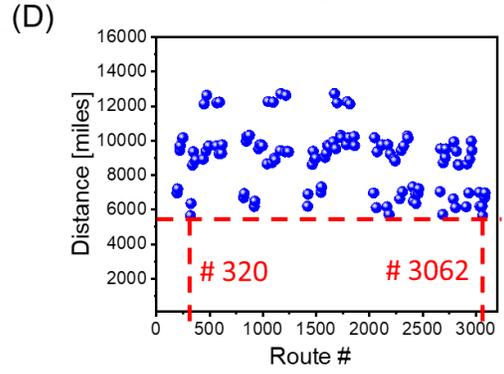

**Figure 5**

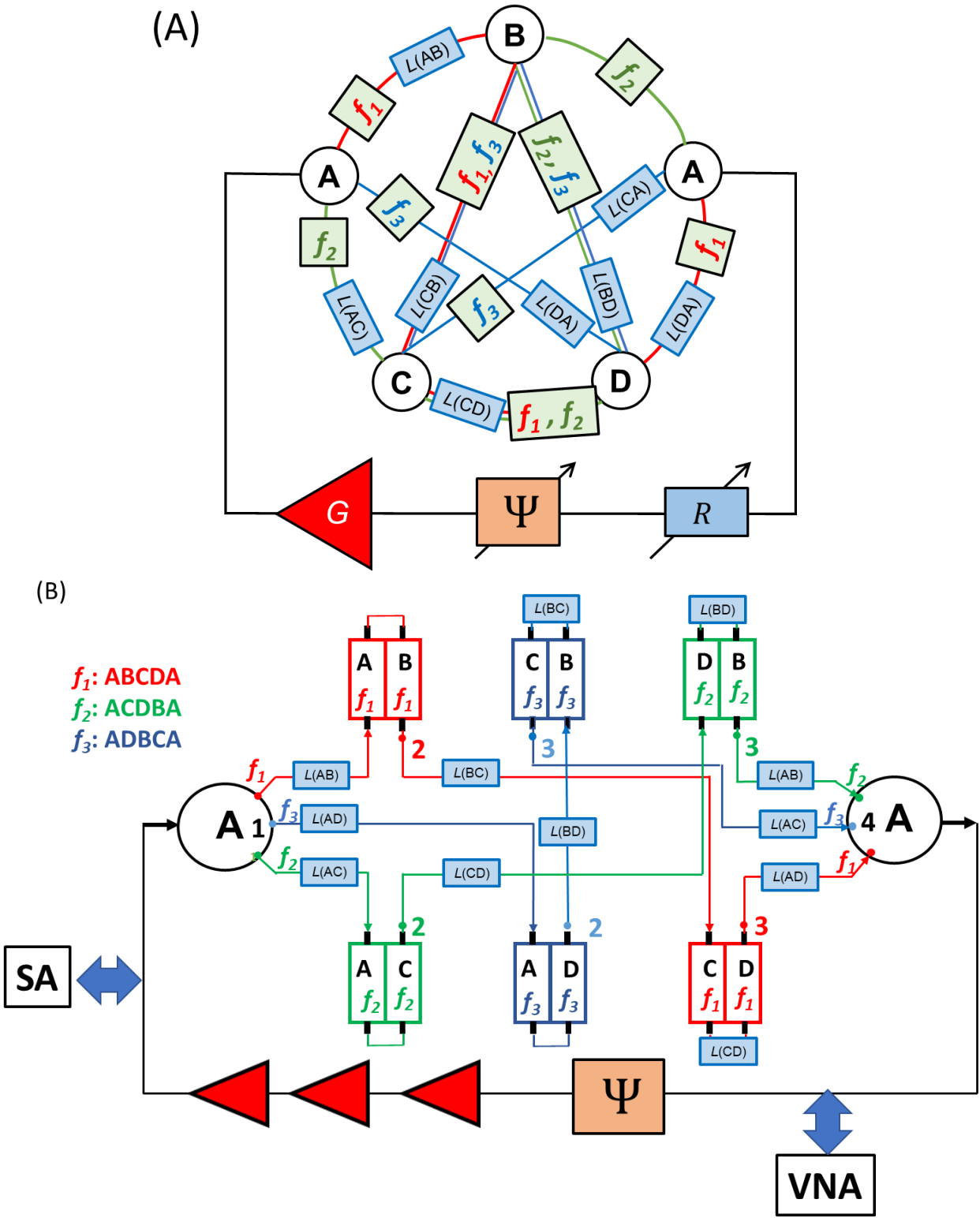

**Figure 6**

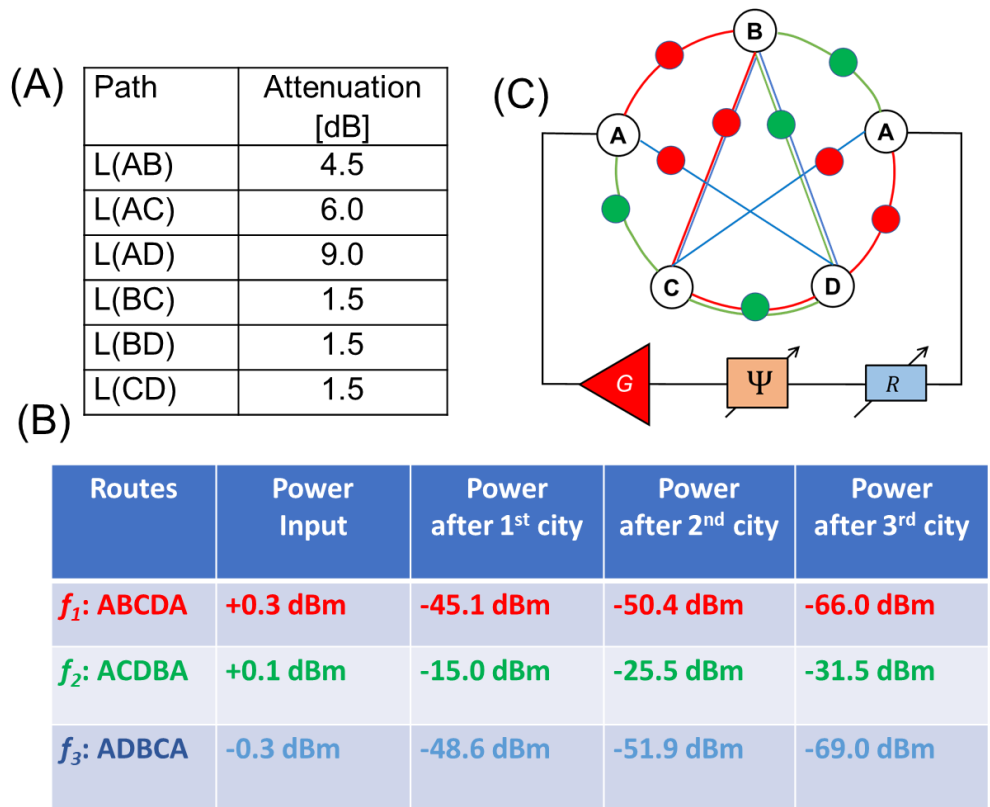

(A)

| Path | Attenuation [dB] |
|---|---|
| L(AB) | 4.5 |
| L(AC) | 6.0 |
| L(AD) | 9.0 |
| L(BC) | 1.5 |
| L(BD) | 1.5 |
| L(CD) | 1.5 |

(B)

| Routes | Power Input | Power after 1st city | Power after 2nd city | Power after 3rd city |
|---|---|---|---|---|
| $f_1$: ABCDA | +0.3 dBm | -45.1 dBm | -50.4 dBm | -66.0 dBm |
| $f_2$: ACDBA | +0.1 dBm | -15.0 dBm | -25.5 dBm | -31.5 dBm |
| $f_3$: ADBCA | -0.3 dBm | -48.6 dBm | -51.9 dBm | -69.0 dBm |

**Figure 7**

(A)

| Path | Attenuation [dB] |
|---|---|
| L(AB) | 9.9 |
| L(AC) | 6.6 |
| L(AD) | 3.0 |
| L(BC) | 1.5 |
| L(BD) | 1.5 |
| L(CD) | 1.5 |

(C)

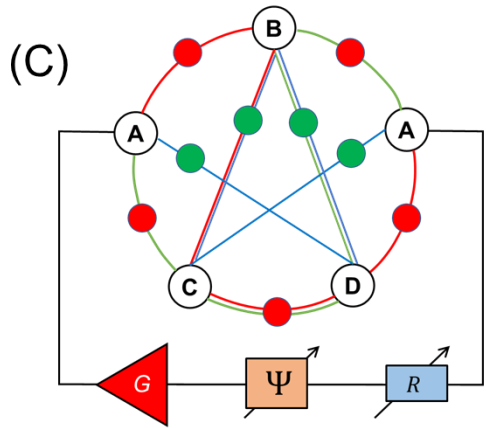

(B)

| Routes | Power Input | Power after 1st city | Power after 2nd city | Power after 3rd city |
|---|---|---|---|---|
| $f_1$: ABCDA | -0.6 dBm | -37.5 dBm | -45.1 dBm | -48.6 dBm |
| $f_2$: ACDBA | -0.9 dBm | -40.0 dBm | -48.3 dBm | -57.0 dBm |
| $f_3$: ADBCA | -1.0 dBm | -10.5 dBm | -18.0 dBm | -24.0 dBm |

**Figure 8**

(A)

| Path  | Attenuation [dB] |
|-------|------------------|
| L(AB) | 6.3              |
| L(AC) | 10.1             |
| L(AD) | 6.6              |
| L(BC) | 1.5              |
| L(BD) | 1.5              |
| L(CD) | 1.5              |

(C) 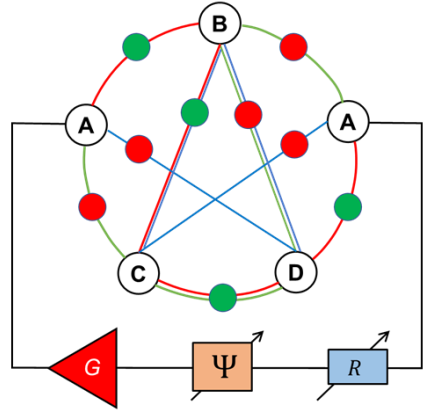

(B)

| Routes | Power Input | Power after 1st city | Power after 2nd city | Power after 3rd city |
|--------|-------------|----------------------|----------------------|----------------------|
| $f_1$: ABCDA | +2.7 dBm | -12.6 dBm | -20.1 dBm | -25.5 dBm |
| $f_2$: ACDBA | +2.6 dBm | -48 dBm   | -51.3 dBm | -57.0 dBm |
| $f_3$: ADBCA | +2.7 dBm | -50.1 dBm | -55.5 dBm | -63.0 dBm |

**Figure 9**

| Component for TSP with $N$ cities | # of parts |
|---|---|
| Phase shifters | $n + 1$ |
| Waveguides | $(n+2)(n-1)/2$ |
| Attenuators | $(n+2)(n-1)/2$ |
| Power sensors | $(n+2)(n-1)/2$ |
| Frequency filters* | $(n+2)(n-1)/2$ [combination of $N$ frequencies per filter] |
| External Broadband Amplifier | 1 |
| External Phase Shifter | 1 |
| External Attenuator | 1 |

**Figure 10**

Supplementary Materials for

**Traveling Salesman Problem solution using Magnonic Combinatorial Device**

| Cities | Distance [miles] |
|---|---|
| Los Angeles – Miami | 2342 |
| Los Angeles – Chicago | 1741 |
| Los Angeles – Washington | 2288 |
| Los Angeles – Las Vegas | 236 |
| Los Angeles – San Francisco | 347 |
| Miami – Chicago | 1189 |
| Miami – Washington | 920 |
| Miami – Las Vegas | 2171 |
| Miami – San Francisco | 2590 |
| Chicago – Washington | 594 |
| Chicago – Las Vegas | 1039 |
| Chicago – San Francisco | 1853 |
| Washington – Las Vegas | 2085 |
| Washington – San Francisco | 2435 |
| Las Vegas – San Francisco | 416 |

Table with city-to-city distances used in numerical simulations. The distances between the cities are taken from the Google Map application.

**YIG-based frequency filter produced by Micro Lambda Wireless, Inc, model MLFD-40540**

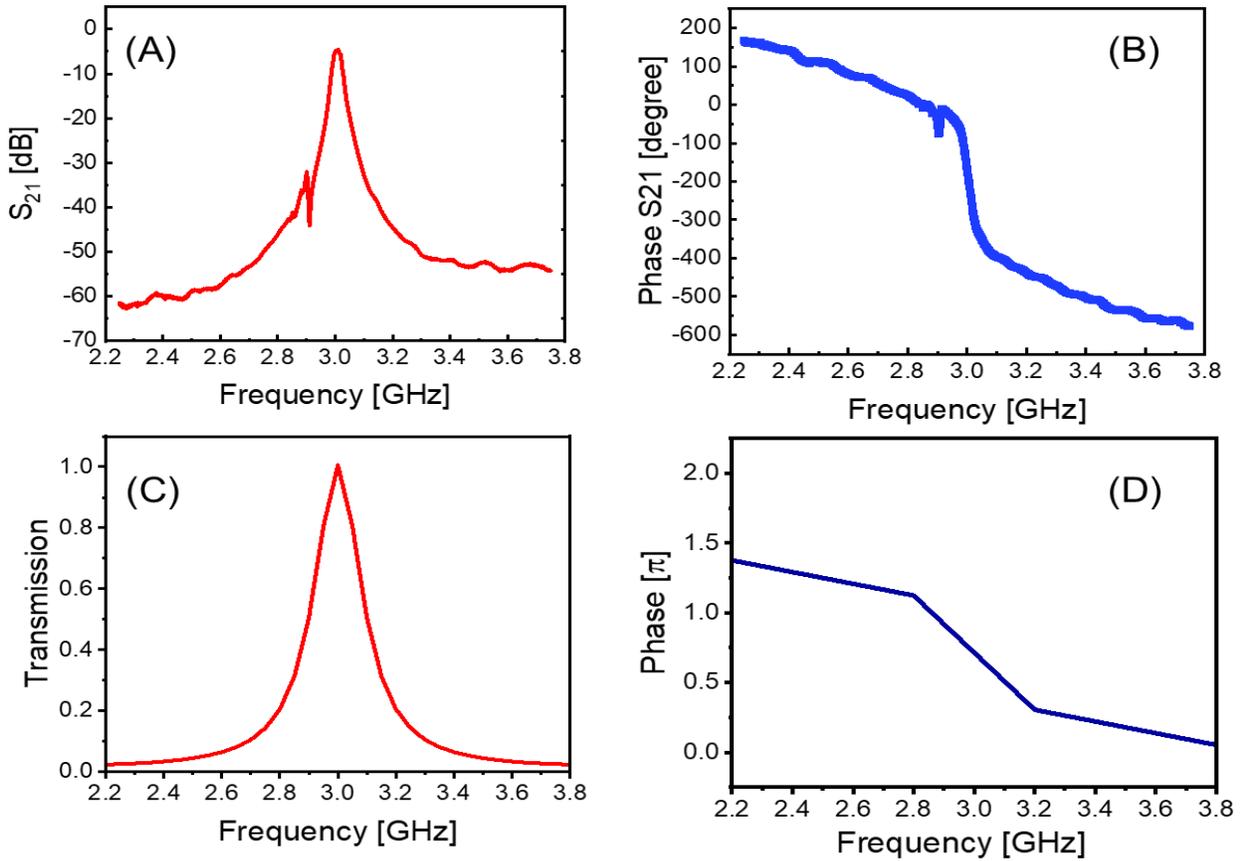

(A) Experimental data: S21 parameter (amplitude) of the commercial filter. (B) Experimental data: S21 parameter (phase shift) of the commercial filter. (C) Results of numerical fitting: transmission of the spin wave element. (D) Results of numerical fitting: phase shift produced by the spin wave element. The data are shown in the frequency range from 2.2 GHz to 3.8 GHz.

Photo of the prototype device

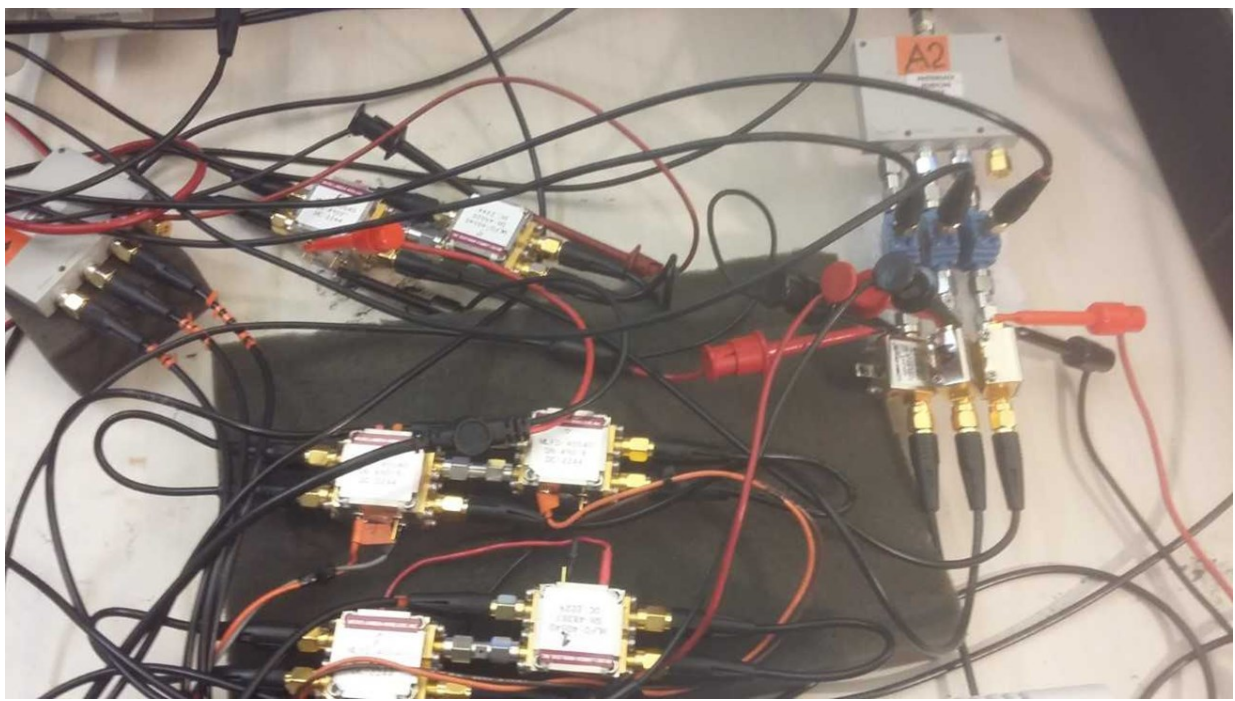

The prototype is built of YIG-based frequency filters produced by Micro Lambda Wireless, Inc, model MLFD-40540 connected by coaxial cables through the set of splitters and combiners.

Experimental data: Transmission of the mesh for the three routes

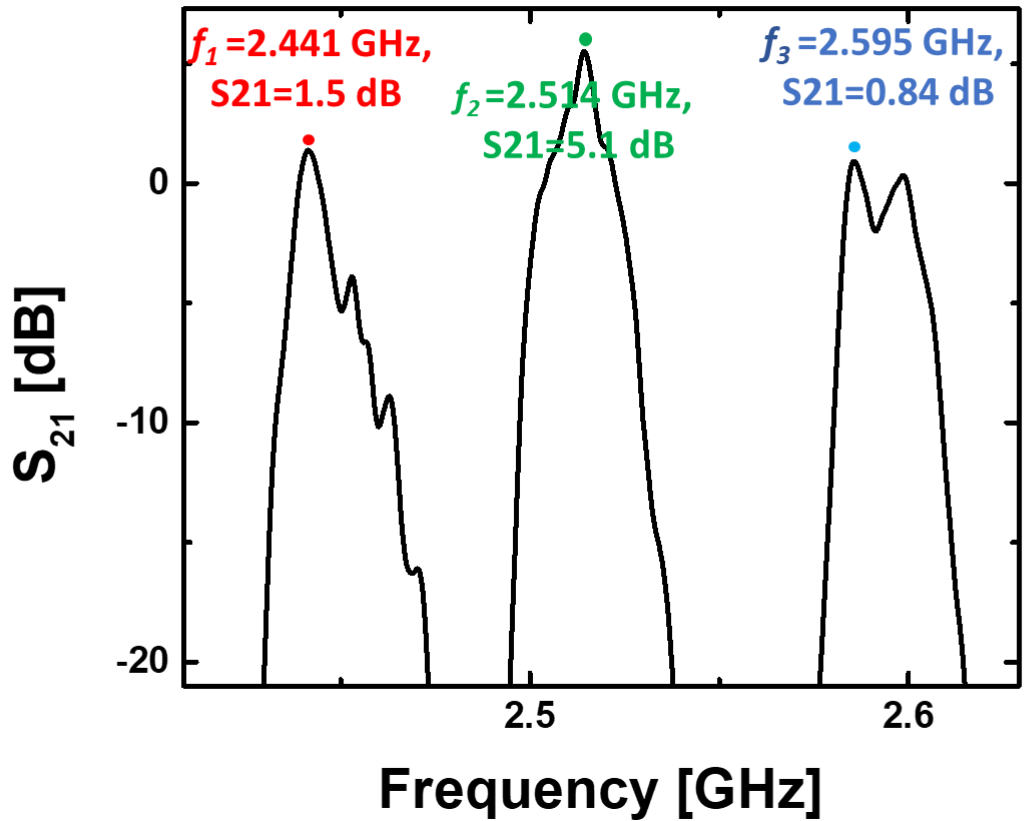

Data are taken with VNA. The mesh is not connected to the electric part (i.e., no amplification)

# Experimental data: Auto-Oscillations in the Active Ring Circuit

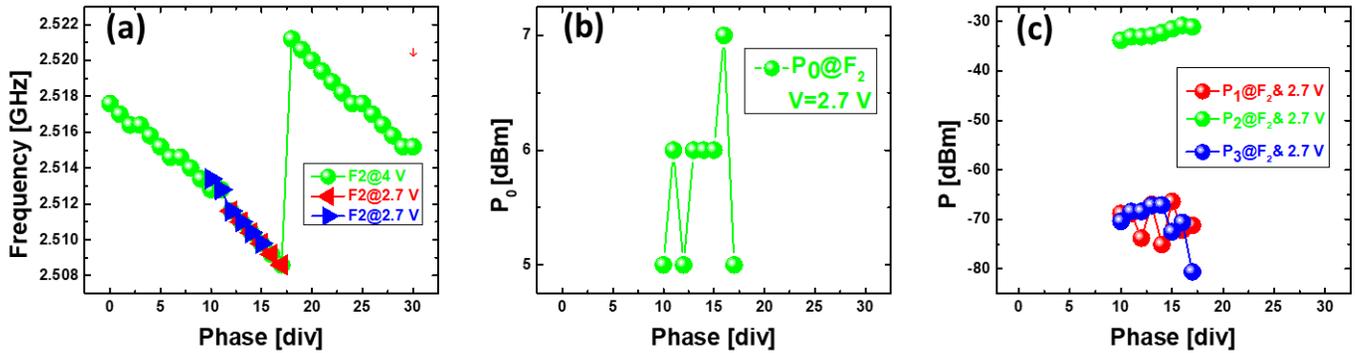

- Magnetic mesh is connected to the electric part. Auto-oscillations are observed for certain levels of amplification and the external phase shifter. The gain in the circuit is set to +13.5 dB.
- (a) Frequencies of the auto-oscillations depending on the external phase shifter. One div is equivalent to $\pi/30$ radians
- (b) Total power in the active ring circuit depending on the external phase.
- (c) Power of the auto-oscillations in different propagation routes.